\documentclass[amssymb,nobibnotes,nofootinbib,aps,prd]{revtex4}

 \usepackage{here}

\usepackage[dvipdfmx]{graphicx}
\usepackage{amssymb,amsfonts,amsmath,bm,xcolor,multirow,cases,empheq,hyperref}
\usepackage{mathrsfs}
\definecolor{darkergreen}{rgb}{0,0.5,0}

\usepackage{enumerate}
\usepackage{ulem}
\usepackage{afterpage}

\hypersetup{%
 setpagesize=false,
 bookmarksnumbered=true,%
 bookmarksopen=true,%
 colorlinks=true,%
 linkcolor=blue,%
 citecolor=red}

\usepackage{tikz}
\usetikzlibrary{positioning,fit}
\usetikzlibrary{calc,intersections,through,backgrounds}
\usetikzlibrary{calc,arrows,decorations.markings,shapes.arrows,patterns, decorations.pathmorphing}
\usetikzlibrary{positioning,matrix,arrows,decorations.pathmorphing,decorations.pathreplacing}
\usetikzlibrary{arrows,matrix,positioning,shapes,arrows}
\usetikzlibrary{shapes.geometric}
\usetikzlibrary{decorations.text}
\usetikzlibrary{arrows.meta}

\tikzset{
  ->-/.style={decoration={markings, mark=at position 0.5 with {\arrow{to}}},
              postaction={decorate}},}
\tikzset{
  -<-/.style={decoration={markings, mark=at position 0.5 with {\arrow{to reversed}}},
              postaction={decorate}},}
              
\tikzset{
  pics/torus/.style n args={3}{
    code = {
      \providecolor{pgffillcolor}{rgb}{1,1,1}
      \begin{scope}[
          yscale=cos(#3),
          outer torus/.style = {draw,line width/.expanded={\the\dimexpr2\pgflinewidth+#2*2},line join=round},
          inner torus/.style = {draw=pgffillcolor,line width={#2*2}}
        ]
        \draw[outer torus] circle(#1);\draw[inner torus] circle(#1);
        \draw[outer torus] (180:#1) arc (180:360:#1);\draw[inner torus,line cap=round] (180:#1) arc (180:360:#1);
      \end{scope}
    }
  }
}

\newcommand{\tikznode}[2]{\relax
    \ifmmode%
    \tikz[remember picture,baseline=(#1.base),inner sep=0pt] \node (#1) {$#2$};
    \else
    \tikz[remember picture,baseline=(#1.base),inner sep=0pt] \node (#1) {#2};%
    \fi
}

\newcommand{\no}{\nonumber}

\newcommand{\cH}{\mathcal H}

\newcommand{\cL}{\mathcal L}
\newcommand{\cM}{\mathcal M}\newcommand{\cN}{\mathcal N}
\newcommand{\cO}{\mathcal O}
\newcommand{\cQ}{\mathcal Q}

\newcommand{\sfA}{\mathsf A}

\newcommand{\la}{\lambda}
\newcommand{\Tr}{{\rm Tr}}

\allowdisplaybreaks

\begin{document}

\begin{flushright}
TTI-MATHPHYS-42\\
OU-HET-1316
\end{flushright}
\vspace*{0.5cm}

\title{
Multi--black holes in  Bertotti--Robinson spacetime
}

\author{Hideo Furugori$^{1}$ }
\email{hideo@toyota-ti.ac.jp}
\author{Jun-ichi Sakamoto$^{2}$ }
\email{jsakamoto@het.phys.sci.osaka-u.ac.jp}
\author{Shinya Tomizawa$^{1}$ }
\email{tomizawa@toyota-ti.ac.jp}
\affiliation{\vspace{3mm}$^{1}$Mathematical Physics Laboratory, Toyota Technological Institute\vspace{2mm}\\Hisakata 2-12-1, Tempaku-ku, Nagoya, Japan 468-8511\vspace{2mm}\\
$^{2}$Department of Physics, The University of Osaka\vspace{2mm}\\
Machikaneyama-Cho 1-1, Toyonaka, Japan 560-0043
\vspace{3mm}}

\begin{abstract}
We construct a new class of exact solutions describing multi-black holes in the Bertotti--Robinson  spacetime, using the monodromy-matrix formalism associated with integrable sigma models. 
Starting from the extremal Reissner--Nordstr\"om black hole in the Bertotti--Robinson background, we derive the corresponding coset and monodromy matrices and show that they are governed by nilpotent algebraic structures. 
This property enables an explicit factorization of the monodromy matrix, allowing for a systematic reconstruction of the underlying gravitational solutions.
We extend this construction to multi-center configurations by introducing multiple poles in the monodromy matrix, leading to Majumdar--Papapetrou--type solutions with Bertotti--Robinson asymptotics. Each center is shown to correspond to a regular extremal black hole with an $\mathrm{AdS}_2 \times S^2$ near-horizon geometry, and the asymptotic end likewise approaches a Bertotti--Robinson geometry.
We further generalize the framework to stationary configurations in the Bertotti--Robinson spacetime, as well as to a broader class of Israel--Wilson--Perj\'es-type solutions, by considering more general nilpotent elements. 
Our results demonstrate that the monodromy-matrix approach provides a powerful and systematic framework for constructing multi-black hole solutions in nontrivial backgrounds, and suggest a promising route toward more general configurations.

\end{abstract}

\date{\today}

\maketitle

\section{Introduction}

Exact solutions describing systems of multiple black holes are of enduring interest in both astrophysics and theoretical physics, as they offer valuable insight into black hole binaries--key sources in gravitational--wave astronomy. However, constructing such solutions remains highly nontrivial, primarily due to the absence of sufficient symmetries and the intrinsically dynamical nature of multi-body gravitational interactions.
Despite these challenges, several classes of static and stationary multi-black hole solutions have been constructed. One of the earliest examples is the solution of Israel and Khan~\cite{Israel1964}, which describes a static, axisymmetric configuration of multiple Schwarzschild black holes aligned along a common axis. In these configurations, gravitational attraction between the black holes necessitates the presence of conical singularities (or struts) along the axis in order to maintain equilibrium.
A rotating generalization of this solution was later constructed by Kramer and Neugebauer~\cite{Kramer1980}, who derived the double Kerr solution representing two interacting rotating black holes. This solution captures the interplay between gravitational attraction and spin-spin repulsion arising from the angular momenta. Nevertheless, conical singularities persist in the region between the black holes, indicating that exact vacuum configurations of static or stationary multi-black hole systems generally involve such singular structures.

\medskip

The existence of electric charge, however, qualitatively alters this picture. A notable example is the Majumdar–Papapetrou solution~\cite{Majumdar:1947eu,Papapetrou}, which provides an exact static multi-black hole solution of the Einstein–Maxwell equations. In this case, gravitational attraction is precisely balanced by electrostatic repulsion, allowing for equilibrium configurations without conical singularities. Israel and Wilson~\cite{Israel:1972vx}, together with Perj\'es~\cite{Perjes:1971gv}, extended this construction to include rotation, obtaining a class of stationary and asymptotically flat solutions within Einstein–Maxwell theory.
It was subsequently shown, however, that these configurations correspond to naked singularities rather than genuine black holes. This suggests that equilibrium configurations of rotating charged black holes with flat asymptotics cannot be realized within Einstein–Maxwell theory alone.
More recently, Teo and Wan~\cite{Teo:2023wfd} constructed a new class of exact, fully regular solutions describing multi-centered spinning black holes in five-dimensional Kaluza--Klein theory. After dimensional reduction, these solutions yield balanced configurations of arbitrarily many dyonic rotating black holes in four-dimensional Einstein--Maxwell--dilaton theory. Each constituent is characterized by independent parameters specifying its mass, spin angular momentum, electric and magnetic charges, and spatial position. 
In the limit where all spin angular momenta vanish, the dilaton field also disappears, and the solution smoothly reduces to the Majumdar--Papapetrou configuration.

\medskip
The examples discussed above concern equilibrium configurations supported by interactions among the black-hole constituents themselves. A qualitatively different situation arises when black holes are embedded in external electromagnetic fields. Two canonical examples of such backgrounds are the Bonnor--Melvin~\cite{Bonnor:1954tis,Melvin:1963qx} and Bertotti--Robinson spacetimes~\cite{Bertotti:1959pf,Robinson:1959ev}, and a variety of black hole solutions in these electromagnetic environments have been constructed~\cite{Ernst:1976mzr,Ernst:1976bsr,Aliev:1989wz,Alekseev:1996fq,Podolsky:2025tle,Ovcharenko:2025cpm,Astorino:2025lih,Alekseev:2025czq,Ovcharenko:2026byw,Astorino:2026kuv}. It is therefore natural to ask whether the no-force mechanism persists under such non-asymptotically flat boundary conditions, and whether it can support configurations with multiple disconnected black-hole horizons.
A suitable framework for addressing this question has recently been developed by Ovcharenko and Podolsk\'y. They constructed a broad class of stationary and axisymmetric exact solutions to the Einstein--Maxwell equations of Petrov type~D with non-aligned electromagnetic fields~\cite{Podolsky:2025tle,Ovcharenko:2025cpm}. A subsequent study identified non-twisting sectors that include several static subfamilies~\cite{Ovcharenko:2026byw}. In particular, one such subfamily describes an accelerating Reissner--Nordstr\"om black hole embedded in a Bertotti--Robinson electromagnetic universe, which we refer to as the RN--BR solution. The Bertotti--Robinson geometry, given by $\mathrm{AdS}_2\times S^2$, provides a natural non-asymptotically flat electromagnetic background for black holes.
The general RN--BR solution may exhibit both an acceleration horizon and conical singularities along the symmetry axis. The extremal limit, however, has a qualitatively different structure: the inner and outer black-hole horizons merge into a degenerate horizon, the acceleration horizon disappears, and the symmetry axis becomes regular.
These features motivate the central question of the present work, namely whether the extremal RN--BR geometry admits a multi-center generalization in which several degenerate horizons coexist within a spacetime with Bertotti--Robinson asymptotics.

\medskip
In Einstein gravity and supergravity, assuming the existence of additional Killing vectors reduced the field equations to those of a two-dimensional integrable coset sigma model~\cite{Maison:1979kx} defined on a conformally flat space. The corresponding action takes the form
\[
S=\int d\rho dz\rho \mathrm{Tr}\left(M^{-1}\partial_m M, M^{-1}\partial^m M\right),
\]
where $M=M(x)$ ( $x=(\rho,z)$: Weyl--Papapetrou coordinates) is a coset matrix on a certain symmetric space $G/H$, which depends on gravity theories and matter contents. For example, it corresponds to $SL(2,{\mathbb R})/SO(2)$ for Einstein gravity, $SU(2,1)/S[U(2)\times U(1)]$ for Einstein--Maxwell theory, $SL(3,{\mathbb R})/SO(2,1)$ for 5D Einstein gravity, $G_{2(2)}/[SL(2,{\mathbb R})\times SL(2,{\mathbb R})]$ for 5D minimal supergravity, $SO(4,4)/[SO(2,2)\times SO(2,2)]$ for $U(1)^3$ supergravity. 
The coset matrix $M(x)$ can be written as $M(x) = V^{\natural}(x)V(x)$ in terms of a certain coset representative $V(x)$.
A powerful solution-generating method in this framework was based on the Breitenlohner--Maison (BM) linear system~\cite{Breitenlohner:1986um}, which can be expressed as a linear equation of  the coset element ${\cal V}(\lambda,x)$. 
This unifies various techniques, including the inverse scattering method and sigma-model transformations such as the Ehlers and Harrison transformations. 
The central object in this framework is the monodromy matrix ${\cal M}(w)= {\cal V}^{\natural}(\lambda,x){\cal V}(\lambda,x)$, a meromorphic function of the spectral parameter $w$ taking values in the Geroch group. 
The coset matrix $M(x)$ can be systematically constructed by performing the factorization of the monodromy matrix,
$
{\cal M}(w) 
= X_-(\lambda,x) M(x) X_+(\lambda,x),
$
where $\natural$ denotes an anti-involution. The matrices $X_+(\lambda,x)$ and $X_-(\lambda,x)$ are defined by
$
X_+(\lambda,x) := V^{-1}(x){\cal V}(\lambda,x), 
X_-(\lambda,x) := X_+^{\natural}(-1/\lambda,x).
$

\medskip
Motivated by these developments, the previous works~\cite{Sakamoto:2025xbq,Sakamoto:2025sjq} constructed monodromy matrices for various five-dimensional black hole solutions and showed that their factorization can reproduce the corresponding geometries. 
The recent work~\cite{Sakamoto:2026cyo}  extended these analyses to extremal black hole solutions in $U(1)^3$ supergravity, focusing on stationary biaxisymmetric configurations over a Gibbons--Hawking base, where for  both BPS and almost-BPS solutions, their monodromy-matrix structures were investigated.
These works showed that, for a broad class of extremal BPS solutions, the monodromy matrices can be factorized even in the presence of higher-order poles by exploiting nilpotent subalgebras of $\mathfrak{so}(4,4)$. 
This then extended this framework to almost-BPS solutions, where the pole structure becomes more intricate. 
In particular, it can be found that higher-order poles disappear precisely when regularity conditions are imposed, demonstrating a direct connection between the analytic structure of the monodromy matrix and the physical regularity of the solution.

\medskip

In this work, we extend the monodromy-matrix construction of asymptotically flat and asymptotically flat Kaluza--Klein extremal black holes in~\cite{Sakamoto:2026cyo} to asymptotically Bertotti--Robinson spacetimes, and obtain a new class of exact multi--extremal black hole solutions with Bertotti--Robinson asymptotics.
Beginning with the extremal Reissner--Nordstr\"om black hole in the Bertotti--Robinson background, we obtain the associated coset and monodromy matrices and demonstrate that they are controlled by nilpotent algebraic structures. 
This structure allows for an explicit factorization of the monodromy matrix, which in turn enables a systematic reconstruction of the corresponding gravitational solutions.
We then extend this construction to multi-center configurations by introducing multiple poles in the monodromy matrix, yielding Majumdar--Papapetrou--type solutions with Bertotti--Robinson asymptotics. 
Each center represents a regular extremal black hole with an $\mathrm{AdS}_2 \times S^2$ near-horizon geometry, and the asymptotic end likewise approaches a Bertotti--Robinson spacetime.
Furthermore, we generalize the framework to include stationary configurations and a wider class of Israel--Wilson--Perj\'es-type solutions by allowing more general nilpotent elements. 
Overall, our results establish the monodromy-matrix approach as a robust and systematic method for constructing multi-black hole solutions in nontrivial backgrounds, and point toward its potential applicability to more general classes of configurations.

\medskip
The remainder of this paper is organized as follows. In Sec.~II, we review the extremal Reissner--Nordstr\"om black hole in the Bertotti--Robinson background and its basic properties. In Sec.~III, we present the sigma-model formulation and introduce the monodromy-matrix framework. In Sec.~IV, we construct the monodromy matrix for the extremal solution and demonstrate its factorization. In Sec.~V, we generalize the construction to multi-center configurations and derive the corresponding solutions. In Sec.~VI, we further extend the framework to a broader class of Israel--Wilson--Perj\'es-type solutions. 
Finally, Sec.~VII is devoted to a summary and discussion of our results.

\section{Reissner--Nordstr\"om Black Hole in Bertotti--Robinson Universe}
\label{Sec. RN--BR}

Recently, Ovcharenko and Podolsk\'y constructed a new class of Petrov type D exact solutions to the Einstein–Maxwell equations with non-aligned electromagnetic field, describing various black holes immersed in an external electromagnetic field~\cite{Ovcharenko:2025cpm}.
In particular, the non-twisting sector contains a family describing an accelerated Reissner--Nordstr\"om (RN) black hole in the Bertotti--Robinson (BR) background~\cite{Ovcharenko:2026byw}.
We refer to this family as the RN--BR spacetime.

\medskip

In this section, we briefly review the RN--BR spacetime. 

\subsection{The Reissner--Nordstr\"om--Bertotti--Robinson spacetime}

We first review the RN--BR spacetime, which is an exact solution of the four-dimensional Einstein--Maxwell theory. 
The action is given by
\begin{align}
S &= \int R_4\star_4 1 - 2\star_4 F\wedge F\,.
\end{align}
Here $R_4$ is the four-dimensional Ricci scalar, $A$ is the Maxwell potential, $F=dA$ is its field strength, and $\star_4$ denotes the Hodge dual operator.
The equations of motion are 
\begin{align}
    &R_{mn}-\frac{1}{2}g_{mn}R_4=2\left(F_{mp}F_{n}{}^{p}-\frac{1}{4}g_{mn}F_{pq}F^{pq}\right)\,,\\
    & d\star_4 F=0\,.
\end{align}

\medskip

The non-twisting sector of the Ovcharenko--Podolsk\'y solution splits into two branches, characterized by $r_0=0$ and $r_0\neq0$. The parameter $r_0$ is associated with the black-hole charge $q$ and is also related to the strength $B$ of the Bertotti--Robinson electromagnetic field.
We focus on the nonzero $r_0$ branch corresponding to the RN--BR spacetime.
In this spacetime, the parameter $r_0$ can be determined by
\begin{align}
     r_{0\pm}
  =
  \frac{M}{I}
  \left(
    \pm\sqrt{1+\frac{Iq^2}{(CM)^2}}-1
  \right)\,,
\end{align}
where $M$ is the mass parameter and $C$ is the conicity parameter representing the periodicity of
azimuthal coordinate
\begin{align}
    \phi\sim\phi+2\pi C\,,
\end{align}
and we have defined 
$I \coloneqq 1-B^2M^2$.
The metric of the RN--BR spacetime can be written as \cite{Ovcharenko:2026byw}
\begin{equation}
  ds^2
  =
  \frac{1}{\Omega^2}
  \left[
    -Q\,dt^2
    +\frac{dr^2}{Q}
    +r^2
    \left(
      \frac{d\theta^2}{P}
      +P\sin^2\theta\,d\phi^2
    \right)
  \right]\,.
\end{equation}
with the metric functions:
\begin{align}
  P(\theta)&= 1-2\alpha_\pm M\cos\theta+B^2M^2\cos^2\theta\,,
  \\
  Q(r)&=\left(
    I-\frac{2M_\pm}{r}+\frac{e^2+g^2}{r^2}\right)
  \Bigl[1+(B^2-\alpha_\pm^2)(r-r_{0\pm})^2
  \Bigr]\,,\label{Q}
  \\
  \Omega^2(r,\theta)&=\left[1-\left(\alpha_\pm r-\frac{qB}{C}\right)\cos\theta
  \right]^2
  +B^2\left[(r-r_{0\pm})^2(P-\cos^2\theta)+2M(r-r_{0\pm})\cos^2\theta
  \right]\,,
\end{align}
and the parameters: 
\begin{align}
  M_\pm=M+I r_{0\pm}
  =
  \pm\sqrt{M^2+I(e^2+g^2)}\,,
  \qquad
  e^2+g^2=\frac{q^2}{C^2}\,,
\end{align}
where $e$ and $g$ denote electric and magnetic charge parameters, respectively.
The acceleration parameter $\alpha_{\pm}$ 
can be written as
\begin{align}
    \alpha_\pm
  =
  \frac{BIq}{CM}
  \left(
    \pm\sqrt{1+\frac{Iq^2}{(CM)^2}}-1
  \right)^{-1}\,.
\end{align}
The absence on the north ($\theta=0$) and south ($\theta=\pi$) components of the symmetry axis requires, respectively,
\begin{align}
    C P(0)=1,
    \qquad
    C P(\pi)=1.
\end{align}
Thus, one may choose
\begin{align}
    C=\frac{1}{1-2\alpha_\pm M+B^2M^2}
\end{align}
to regularize the axis at $\theta=0$ only, while generically, the axis at
$\theta=\pi$ then contains conical singularities. 
Both components of the axis are simultaneously free of conical singularities if and only if
\begin{align}
    \alpha_\pm M=0,
    \qquad
    C=\frac{1}{1+B^2M^2}.
\end{align}

\medskip

The parameters $\alpha_\pm$ and $r_{0\pm}$ are not independent and simply related  through
\begin{equation}
  C r_{0\pm}\alpha_\pm=qB.
\end{equation}
This relation shows that the acceleration is tied to the interaction
between the black-hole charge and the external Bertotti--Robinson electromagnetic field.
We note that in order to describe a black hole spacetime rather than a naked singularity, the following condition must be satisfied:
\begin{align}
    |B|\sqrt{e^2+g^2}\le 1\,.
\end{align}
In this case, the inner and outer black hole horizons are located at
\begin{align}
    r_{\mathrm{b}-}=r_{0\pm}\,,\quad 
    r_{\mathrm{b}+}=r_{0\pm} + \frac{2M}{I}\,.
\end{align}
We also note that when $\alpha_\pm > B$, an acceleration horizon appears at the root of the second bracket in the $Q$ of eq.~(\ref{Q}).

\subsection{Extremal Reissner–Nordstr\"om -- Bertotti-Robinson black hole}

The extremal limit of RN--BR black hole is obtained by taking the following limit:
\begin{equation}
  M\to0,
  \quad  
  \text{with } q=\text{fixed}\,.
\end{equation}
 Here and in what follows, the branch sign is chosen to coincide with the sign of $q$.
For definiteness, we assume $q>0$ and select the $+$ branch.
We then define $m\coloneqq q$.
In this limit, one has
\begin{equation}
  I\to 1\,,
  \quad
  r_{0+}\to q=m\,,\quad 
  M_+ \to q = m
\,,\quad 
\alpha_+ \to B\,.
\end{equation}

We also obtain 
\begin{equation}
C=1\,, 
\end{equation}
indicating that there is no conical singularity.

\medskip
The metric functions reduce to
\begin{align}
  P=1\,,\qquad
  Q=
  \left(
    1-\frac{m}{r}
  \right)^2\,,\qquad
  \Omega^2
  =1-2B(r-m)\cos\theta+B^2(r-m)^2\,.
\end{align}
The resulting metric becomes
\begin{align}
ds^2 = \frac{1}{\Omega^2} \left[
- \left(1-\frac{m}{r}\right)^2\, dt^2 +\left(1-\frac{m}{r}\right)^{-2} dr^2+ r^2 \left( d\theta^2 + \sin^2\theta \, d\phi^2 \right)
\right]\,.
\end{align}
Thus, the inner and outer horizons coincide at
\begin{align}
   r_{\mathrm{b}+}= r_{\mathrm{b}-}=m\,.
\end{align}
Note that there is no acceleration horizon in this spacetime. 
The electromagnetic potential can be written by 
\begin{align}\label{gauge-exrn}
A = -\sin\gamma \frac{r-m}{r\Omega}dt+\cos\gamma \Bigg[
\frac{r(B (r-m)-\cos\theta)}{\Omega}-\frac{\Omega-1}{B}
\Bigg]\,d\phi\,.
\end{align}
The solution describes an extremal Reissner--Nordstr\"om black hole
immersed in Bertotti--Robinson electromagnetic field as desired.

\medskip

When $m=0$, the solution reduces to the Bertotti--Robinson universe in accelerating coordinates.
The metric is given by
\begin{align}
ds^2 &= \frac{1}{(1-Br\cos \theta)^2 +B^2r^2 \sin ^2 \theta} \left[
-dt^2 + dr^2 + r^2 \left(d\theta^2 + \sin ^2 \theta d \phi^2\right)
\right]\,.
\end{align}
This conformally flat form of the metric can be transformed to the standard form with AdS$_2\times S^2$. 
Indeed, by a coordinate transformation $(t,r,\theta)\mapsto (T,R,\Theta)$ given by
\begin{align}
T=Bt\,,\quad 
    R^2 = \Omega^2\,,\quad 
    \sin\Theta = \frac{Br}{\Omega} \sin\theta\,,\quad
    \cos \Theta = \frac{Br\cos \theta-1}{R}\,,
\end{align}
the metric is given by
\begin{align}
    ds^2 = \frac{1}{B^2}\left[
    \frac{-dT^2+dR^2}{R^2}+d\Theta^2 +\sin ^2 \Theta d\phi^2
    \right]\,,
\end{align}
representing the product of AdS$_2$ with radius $1/B$ and $S^2$ with the same radius $1/B$.

\section{Sigma model description}

In this section, we give the 2D integrable sigma model description of the stationary axisymmetric four-dimensional black hole solutions in the 4D Einstein--Maxwell theory by following \cite{Chakrabarty:2014ora}.

\subsection{Sigma model description from Einstein--Maxwell theory}

We suppose stationary axisymmetric four-dimensional black hole solutions in the 4D Einstein--Maxwell theory.
The metric and gauge field are written as
\begin{align}\label{ansatz}
\begin{split}
ds_4^2 &= -e^{-\phi}(dt+\omega)^2 + e^{\phi}ds_3^2\,, \\
A &= \chi_e\,dt + \tilde A\,.
\end{split}
\end{align}
The scalar $\phi$ measures the norm of the timelike Killing vector, $\omega$ is the Kaluza--Klein one-form in three dimensions, $\chi_e$ is the electric potential, and $\tilde A$ is the three-dimensional part of the gauge field. All fields are taken to be independent of $t$ and $\phi$. 

\medskip

To obtain the sigma model description, we first perform a dimensional reduction along a timelike Killing direction $\partial_t$.
After the reduction, we introduce two field strengths as
\begin{align}\label{F-tF}
F &= d\omega\,, \qquad \tilde F = d\tilde A - d\chi_e\wedge \omega\,.
\end{align}
The shifted field strength $\tilde F$ is constructed to be gauge covariant.
In three dimensions, these field strengths can be mapped to the scalar fields $\chi_m$ and $\psi$ through the duality relations
\begin{align}
\tilde F &= -e^{\phi}\star_3 d\chi_m\,, \label{fomega-dual1}\\
F &= e^{2\phi}\star_3\left(2\chi_m d\chi_e -2\chi_e d\chi_m + \sqrt{2}\,d\psi\right)\,.\label{fomega-dual}
\end{align}
After this dualization, the bosonic degrees of freedom of the stationary Einstein--Maxwell system are described entirely by the four scalar fields $\{\phi,\chi_e,\chi_m,\psi\}$.

\medskip

These four scalars $\{\phi,\chi_e,\chi_m,\psi\}$ parametrize the symmetric coset
\begin{align}
    \frac{SU(2,1)}{SL(2,\mathbb{R})\times U(1)}\,.
\end{align}
Using the notation for $SU(2,1)$ summarized in appendix \ref{sec:notation-su21}, we choose a coset representative $V \in SU(2,1)$ as
\begin{align}
V &= \exp\left[\frac{1}{2}\phi h_2\right]
\exp\left[\sqrt{2}\chi_e(e_1+e_2)
+ \sqrt{2}\chi_m\,i(e_2-e_1)
+ \sqrt{2}\psi\,(ie_3)\right]\,.
\end{align}
The representative $V$ itself depends on the choice of local coset gauge. It is therefore convenient to introduce the gauge invariant coset matrix
\begin{align}\label{def-coset}
M &= V^\natural V\,,
\end{align}
where $\natural$ is a certain transposition operator defined in (\ref{transpose}).
By definition, the coset matrix $M$ is invariant under the gauge transformation $V(x)\mapsto h(x) V(x)$ for $h(x) \in SL(2,\mathbb{R})\times U(1)$.

\medskip

The resulting 3D system reduces to the action of a 3D symmetric coset sigma model coupled to 3D Einstein gravity
\begin{align}
S_{3}&=\int  R_3\star_3 1
- \frac{1}{4}\Tr\left(\star_3(M^{-1}dM)\wedge(M^{-1}dM)\right)\,.
\end{align}
In this work, we are mostly interested in axisymmetric solutions.
For axisymmetric solutions, the three-dimensional base metric can be written in Weyl--Papapetrou coordinates as
\begin{align}
    ds_3^2=e^{2\nu}(d\rho^2+dz^2)+\rho^2 d\phi^2\,.
\end{align}
Here $e^{2\nu}$ denotes the conformal factor. Throughout this paper, we consider gravitational solutions with a trivial conformal factor,
\begin{align}
e^{2\nu}=1\,,
\end{align}
or equivalently, solutions for which the 3D base space is flat. This is a typical feature of extremal black holes.

\medskip

Finally, we further reduce the system along the angular direction $\phi$. This gives the 2D action
\begin{align}\label{2dsystem-action}
S_{2}&=\int \rho \Bigl[R_2\star_2 1
- \frac{1}{4}\Tr\left(\star_2(M^{-1}dM)\wedge(M^{-1}dM)\right)\Bigr]\,.
\end{align}
The equations of motion for the sigma-model sector are
\begin{align}\label{sigma-eom}
    \partial_{\rho}(\rho \partial_{\rho}MM^{-1})+\partial_{z}(\rho \partial_{z}MM^{-1})=0\,.
\end{align}
Remarkably, this coordinate-dependent two-dimensional symmetric coset sigma model is classically integrable \cite{Belinsky:1971nt,Belinsky:1979mh}. Its integrable structure makes it possible to construct exact black hole solutions as soliton solutions by using standard solution-generating techniques.

\subsection{Monodromy matrix}

The equations of motion (\ref{sigma-eom}) can be equivalently reformulated as the auxiliary linear problem, usually referred to as the BM linear system,
\begin{align}\label{lin-pro}
    \partial_{\mu}\Psi(\la,z,\rho)=\cL_{\mu}(\la,z,\rho)\Psi(\la,z,\rho)\qquad \mu=z,\rho\,. 
\end{align}
Here, $\cL$ denotes an on-shell flat Lax connection depending on the spectral parameter $\la$. For its explicit form, see, for example, \cite{Sakamoto:2025jtn}. We choose the normalization of $\la$ such that, at $\la=0$ the Lax connection reduces to the right-invariant current
\begin{align}
    \cL_{\mu}(\la=0,z,\rho)=\partial_{\mu}V(z,\rho)V^{-1}(z,\rho)\,.
\end{align}
A characteristic feature of the 2D sigma model in (\ref{2dsystem-action}) is its explicit dependence on the spatial coordinate $\rho$. As a consequence, the spectral parameter $\la$ is not a constant, but rather becomes a variable spectral parameter depending on $\rho$ and $z$ via the algebraic relation 
\begin{align}\label{r-alg}
    \frac{1}{\la}-\la=\frac{2}{\rho}(w-z)\,.
\end{align}
The behavior of the Lax connection at $\la=0$ determines the boundary condition for the auxiliary linear problem (\ref{lin-pro}). Since $\cL_{\mu}(0,z,\rho)=\partial_{\mu}V(z,\rho)V^{-1}(z,\rho)$, the auxiliary linear problem at $\la=0$ is solved by $\Psi(0,z,\rho)=V(z,\rho)$, up to a constant right multiplication. We therefore impose the normalization condition
\begin{align}
\Psi(\la,z,\rho)=V(z,\rho)+\sum_{n=1}^{\infty}\la^{n}\Psi^{(n)}(z,\rho)\,,
\end{align}
so that the original sigma model solution is recovered from $\Psi(\la,z,\rho)$ at $\la=0$.

\medskip

There are several methods for solving the auxiliary linear problem (\ref{lin-pro}). In this work, we adopt an approach based on the monodromy matrix \cite{Breitenlohner:1986um,Chakrabarty:2014ora,Katsimpouri:2012ky,Katsimpouri:2013wka,Katsimpouri:2014ara}. In analogy with the coset matrix, which is constructed from the coset representative as a gauge-invariant combination, the monodromy matrix is defined as a gauge-invariant combination of the auxiliary wave function $\Psi$. More explicitly, using the involution $\natural$ associated with the coset structure, we introduce
\begin{align}
    \cM(w)=\Psi(-1/\la,z,\rho)^{\natural}\Psi(\la,z,\rho)\,.
\end{align}
By definition, it satisfies the algebraic relations
\begin{align}\label{m-con}
\cM^{\natural}=\cM\,, \qquad {\rm det}\,\cM=1\,.
\end{align}
It is noted that the monodromy matrix is independent of the coordinates and is then a matrix-valued meromorphic function of the complex spectral parameter $w\in\mathbb{C}$.
For further details of the monodromy matrix, we refer the reader to the existing literature, e.g. \cite{Breitenlohner:1986um,Chakrabarty:2014ora,Katsimpouri:2012ky,Katsimpouri:2013wka,Sakamoto:2025jtn}.

\medskip

To obtain a solution to the linear problem (\ref{lin-pro}), one first specifies a monodromy matrix. The corresponding classical solution can then be constructed by solving a factorization problem in the variable spectral parameter.
More precisely, after substituting $w=w(\lambda,\rho,z)$ determined by the algebraic relation (\ref{r-alg}), we factorize the monodromy matrix as
\begin{align}\label{mm-fac}
    \cM(w(\la,z,\rho))=X_-(\la,z,\rho)M(z,\rho)X_+(\la,z,\rho)\,,
\end{align}
where the matrix-valued functions $X_+(\la,z,\rho)$ and $X_-(\la,z,\rho)=X_+^{\natural}(-1/\la,z,\rho)$ are required to satisfy the boundary condition
\begin{align}\label{Xpm-bc}
    X_+(0,z,\rho)=1_{3\times 3}\,.
\end{align}
The symbol $w(\la,z,\rho)$ in the left-hand side is to remind us that, when evaluating $\cM$, the variable $w$ is to be regarded as a function of $\la$, $z$, and $\rho$, obtained from a chosen branch of the algebraic relation (\ref{r-alg})
\begin{align}\label{la-w}
    \la=\la(w;z,\rho)=\frac{1}{\rho}\left[(z-w)+ \sqrt{(z-w)^2+\rho^2}\right]\,.
\end{align}
Thus, factorizing a given monodromy matrix allows us to construct the coset matrix $M(z,\rho)$ that solves the equations of motion of the sigma model.

\medskip

However, at present there is no general framework for systematically determining monodromy matrices corresponding to physically meaningful gravitational solutions. On the other hand, for known solutions, it is possible to construct the associated monodromy matrix by taking an appropriate limit of the corresponding coset matrix $M(z,\rho)$ in the limit $\rho \to 0^+$ in a region where $z$ is sufficiently negative:
\begin{align}\label{monodromy-formula}
    \cM(w)=\lim_{\rho \to 0^+}M(z=w,\rho)\qquad \text{for}\quad z<-R\,.
\end{align}
Here, $R$ is chosen as the radius of a semicircle in the upper half-plane $(z,\rho)$ that encloses all finite rods\footnote{In practice, the limit (\ref{monodromy-formula}) is evaluated for $z<-\infty$.}.

\medskip

From previous studies of monodromy matrices for various non-extremal black holes \cite{Chakrabarty:2014ora,Katsimpouri:2012ky,Katsimpouri:2013wka,Katsimpouri:2014ara,Sakamoto:2025jtn,Sakamoto:2025xbq,Sakamoto:2025sjq}, it has been observed that the corresponding monodromy matrices have at most simple poles in $w$.
The positions of simple poles are precisely identified with the locations of the corner points of the rod structure (for the details of the rod structure, see Refs.~\cite{Harmark:2004rm,Hollands:2007aj}).
On the other hand, for extremal black holes, these poles collide, and the monodromy matrix develops double poles:
\begin{align}
    \cM_{\text{extremal}}(w)=Y+\sum_{i=1}^{N}\frac{A_i^{(1)}}{w-w_i}+\sum_{i=1}^{N}\frac{A_i^{(2)}}{(w-w_i)^2}\,,
\end{align}
where the constant matrix $Y$ encodes the asymptotic spacetime structure, $N$ is equal to the number of corner points in the rod structure.
The presence of higher-order poles in the monodromy matrix prevents a straightforward application of the standard procedure developed in \cite{Breitenlohner:1986um,Chakrabarty:2014ora,Katsimpouri:2012ky,Katsimpouri:2013wka,Katsimpouri:2014ara}. 
Fortunately, however, for many extremal black holes the residue matrices are nilpotent, and the monodromy matrix can be reorganized into a single exponential of nilpotent matrices. This property often makes it possible to carry out the factorization by simple algebraic manipulations \cite{Sakamoto:2025sjq,Sakamoto:2026cyo}. Indeed, as we will see below, the monodromy matrix corresponding to the extremal RN--BR black hole can also be written as the exponential of two nilpotent matrices. Moreover, by replacing these nilpotent matrices with a more general collection of nilpotent matrices, one can straightforwardly construct its multi-center generalization, which eventually leads to the Israel-Wilson-Perj\'es solution in the Bertotti--Robinson spacetime.
By contrast, a systematic way of implementing such multi-center generalizations within the inverse scattering method based on the Belinsky-Zakharov linear system has not yet been developed. This constitutes one of the advantages of the present approach.

\section{Monodromy matrix for Extremal RN--BR black hole}

We consider the monodromy matrix for the extremal RN--BR black hole. We find that the monodromy matrix is characterized by nilpotent residue matrices. This property enables us to factorize the monodromy matrix in a straightforward manner.
Using this property of the monodromy matrix, we construct a multi-center extension of the extremal RN--BR black hole.

\subsection{Coset matrix}

We first derive the coset matrix corresponding to the extremal RN--BR black hole. From the ansatz (\ref{ansatz}), and by solving the Hodge duality relations (\ref{fomega-dual1}) and (\ref{fomega-dual}), the four scalar fields are obtained as 
\begin{align}
    e^{\phi}&=\frac{r^2 \Omega^2}{(r-m)^2}\,,\qquad
    \chi_e=\frac{\sin \gamma  (m-r)}{r\Omega}\,,\qquad
    \chi_m=-\frac{\cos \gamma  (m-r)}{r\Omega}\,,\qquad
    \psi=0\,.
\end{align}
One can check that 
\begin{align}
    \chi_m d\chi_e -\chi_e d\chi_m=0\,.
\end{align}
Since $\omega=0$, this relation indicates that $\psi=0$ follows from the duality relation (\ref{fomega-dual}).
Substituting these scalar fields into (\ref{def-coset}) gives the coset matrix
\begin{align}\label{ex-coset}
    M(r,\theta)=\left(
\begin{array}{ccc}
 \frac{r^2 \Omega^2}{(r-m)^2} & \frac{i \sqrt{2} r \Omega e^{i\gamma}}{r-m} & 1 \\
 \frac{i \sqrt{2} r \Omega e^{-i\gamma}}{r-m} & -1 & 0 \\
 1 & 0 & 0 \\
\end{array}
\right)\,.
\end{align}
In contrast to asymptotically flat extremal RN black hole solutions, the present solution yields a divergent asymptotic coset matrix, since $\Omega^2\sim \cO(r^2)$ at large $r$ and hence some components diverge as $r\to\infty$ for generic values of $\theta$.

\medskip

In terms of the Weyl--Papapetrou coordinate system, the 3D metric takes the 3D flat metric
\begin{align}
    ds_3^2&=\frac{1}{\Omega^4}(dr^2+(r-m)^2(d\theta^2+\sin^2\theta d\phi^2))\no\\
    &=d\rho^2+dz^2+\rho^2d\phi^2\,.
\end{align}
Then, the conformal factor $e^{2\nu}$ is trivial,
\begin{align}
    e^{2\nu}=1\,.
\end{align}
This is a feature shared by many extremal black holes, and it has been observed that, in such cases, the coset matrix is often characterized by nilpotent matrices\footnote{Conversely, when $M=\exp(A)$ with $A$ belonging to a nilpotent subalgebra of $\mathfrak{g}$, the conformal factor $e^{2\nu}$ is always trivial \cite{Gaiotto:2007ag}.}.
In what follows, we show that, the coset matrix is indeed characterized by nilpotent matrices.

\subsection{Monodromy matrix}

Next, we derive the corresponding monodromy matrix.
Substituting the coset matrix (\ref{ex-coset}) into the formula (\ref{monodromy-formula}) gives
\begin{align}
    \cM(w)
    &=-\kappa\left(1+\frac{\sfA_1^{(1)}}{w-w_1}+\frac{\sfA_2^{(1)}}{w-w_2}+\frac{\sfA_1^{(2)}}{(w-w_1)^2}+\frac{\sfA_2^{(2)}}{(w-w_2)^2}\right)\,.
\end{align}
Here $\kappa$, defined in (\ref{kappa-rep}), is the constant matrix that characterizes the asymptotic behavior of the fields in the corresponding gravitational solution. The poles are located at
\begin{align}
    w_1=0\,,\qquad w_2=-\frac{1}{B}\,.
\end{align}
The residue matrices are
\begin{align}
\begin{split}
   \sfA_1^{(1)}&=\begin{pmatrix}
0&0&0\\
\sqrt{2} m i\,e^{-i\gamma}&0&0\\
2m& -\sqrt{2}m i\,e^{i\gamma}&0
\end{pmatrix}\,,\qquad
    \sfA_2^{(1)}=\begin{pmatrix}
0&0&0\\
\frac{\sqrt{2} i\,e^{-i\gamma}}{B}&0&0\\
-2m&-\frac{\sqrt{2}i\,e^{i\gamma}}{B}&0
\end{pmatrix}\,,\\
\sfA_1^{(2)}&=\frac{1}{2}( \sfA_1^{(1)})^2\,,\qquad 
\sfA_2^{(2)}=\frac{1}{2}( \sfA_2^{(1)})^2\,.
\end{split}
\end{align}
These matrices are nilpotent
\begin{align}
    \left(\sfA_i^{(1)}\right)^3=0\,.
\end{align}
Then, the monodromy matrix can be rewritten as
\begin{align}
    \cM(w)&=-\kappa\exp\left(\frac{\tilde{\sfA}^{(1)}_1}{w-w_1}+\frac{\tilde{\sfA}^{(1)}_2}{w-w_2}\right)\,,\label{sing-ex-RN}
\end{align}
where we introduced the modified residue matrices by
\begin{align}
\tilde{\sfA}^{(1)}_1&=\sfA_1^{(1)}-\frac{1}{2}\frac{1}{w_1-w_2}\{\sfA_1^{(1)},\sfA_2^{(1)}\}\,,\\
\tilde{\sfA}^{(1)}_2&=\sfA_2^{(1)}+\frac{1}{2}\frac{1}{w_1-w_2}\{\sfA_1^{(1)},\sfA_2^{(1)}\}\,,\\
    \{\sfA_1^{(1)},\sfA_2^{(1)}\}&=\sfA_1^{(1)}\sfA_2^{(1)}+\sfA_2^{(1)}\sfA_1^{(1)}\,.
\end{align}
The modified residue matrices $\tilde{\sfA}^{(1)}_1$ and $\tilde{\sfA}^{(1)}_2$ take forms that depend only on the mass and the electric field, respectively,
\begin{align}
   \tilde{\sfA}^{(1)}_1&=m\,\mathfrak{n}(\gamma)\,,\qquad 
    \tilde{\sfA}^{(1)}_2=\frac{1}{B}\mathfrak{n}(\gamma)\,,\qquad 
\mathfrak{n}(\gamma)=\begin{pmatrix}
0&0&0\\
\sqrt{2} i e^{-i\gamma}&0&0\\
0&-\sqrt{2}i e^{i\gamma}&0
\end{pmatrix}\,, \qquad \mathfrak{n}(\gamma)^3=0\,.
\end{align}
Since these two matrices commute $ [\tilde{\sfA}_1^{(1)},\tilde{\sfA}_2^{(1)}]=0$, it is easy to perform the factorization of the monodromy matrix.

\medskip

For completeness, we examine whether, in the limit $B\to 0$, the monodromy matrix (\ref{sing-ex-RN}) reduces to that corresponding to an asymptotically flat extremal RN black hole. In this limit, the second pole $w=w_2=-B^{-1}$ is pushed to infinity in the $w$-plane. Since the corresponding residue matrix is also proportional to $B^{-1}$, the exponent associated with this pole approaches a finite, nontrivial constant matrix rather than the identity matrix,
\begin{align}
    \lim_{B\to0}\exp\left(\frac{\tilde{\sfA}^{(1)}_2}{w-w_2}\right)=\begin{pmatrix}
1&0&0\\
\sqrt{2} i e^{-i\gamma}&1&0\\
1&-\sqrt{2}i e^{i\gamma}&1
\end{pmatrix}\,.
\end{align}
As a result, the constant matrix $-\kappa$ characterizing the asymptotic structure of the gravitational solution is modified, and the monodromy matrix (\ref{sing-ex-RN}) reduces to the following form:
\begin{align}
    \lim_{B\to0}\cM(w)&=Y\exp\left(\frac{\tilde{\sfA}^{(1)}_1}{w-w_1}\right)\,,\qquad Y=\begin{pmatrix}
1&-\sqrt{2}i e^{i\gamma}&1\\
\sqrt{2} i e^{-i\gamma}&-1&0\\
1&0&0
\end{pmatrix}\label{sing-ex-RN-flat}
\end{align}
This can be verified to be precisely the monodromy matrix corresponding to an asymptotically flat extremal RN black hole \footnote{The constant matrix $Y$ in (\ref{sing-ex-RN-flat}) is different from that associated with the 4D asymptotically flat Kerr-Newman black hole in \cite{Chakrabarty:2014ora}, for which $Y$ is the identity matrix.
This is because the asymptotic matrix $Y$ depends on the choice of gauge for the gauge field at spatial infinity. Indeed, in \cite{Chakrabarty:2014ora}, the gauge field is chosen to vanish at spatial infinity, whereas in our gauge choice (\ref{gauge-exrn}) it approaches a constant value.}.

\subsubsection*{Factorization of monodromy matrix}

As in our previous works \cite{Sakamoto:2026cyo,Sakamoto:2025sjq}, the monodromy matrix can be readily factorized into the form (\ref{mm-fac}), and the coset matrix obtained in this way reproduces the original coset matrix $M(z,\rho)$ in (\ref{ex-coset}). To this end, we rewrite the simple poles in the $w$-plane as
\begin{align}
    \frac{1}{w-w_j}&=\ell_j+\nu_j+r_j\,,\qquad \ell_j=\frac{\nu_j\lambda_j}{\lambda-\lambda_j}\,,\qquad r_j= -\frac{\nu_j\lambda\lambda_j}{1+\lambda\lambda_j}\,,
 \label{eq:pole-splitting}
\end{align}
where we used (\ref{la-w}) and introduced
\begin{align}\label{eq:r-l}
    \la_i&=\frac{1}{\rho}\left(z-w_i+\sqrt{(z-w_i)^2+\rho^2}\right)\,,
\end{align}
The spectral parameter dependent functions are mapped into one another under the transformation $\lambda\to -1/\lambda$:
\begin{align}
     r_j(-1/\lambda)=\ell_j(\lambda)\,.
\end{align}
It is noted that the scalar function $\nu_i$ is a harmonic function
\begin{align}
 \nu_j=-\frac{1}{R_j}\,,\qquad  R_j=\sqrt{(z-w_j)^2+\rho^2}\,.
 \label{eq:nu-A}
\end{align}
Then, the monodromy matrix can be rewritten as
\begin{align}
    \cM(w)&=-\kappa\exp\left(\frac{\tilde{\sfA}^{(1)}_1}{w-w_1}+\frac{\tilde{\sfA}^{(1)}_2}{w-w_2}\right)\no\\
    &=-\kappa\exp\left(\sum_{i=1}^{2}\ell_i\tilde{\sfA}^{(1)}_i\right)\exp\left(\sum_{i=1}^{2}\nu_i \tilde{\sfA}^{(1)}_i\right)\exp\left(\sum_{i=1}^{2}r_i\tilde{\sfA}^{(1)}_i\right)\no\\
    &=\exp\left(\sum_{i=1}^{2}\ell_i\left(\tilde{\sfA}^{(1)}_i\right)^{\natural}\right)\biggl[-\kappa\exp\left(\sum_{i=1}^{2}\nu_i \tilde{\sfA}^{(1)}_i\right)\biggr]\exp\left(\sum_{i=1}^{2}r_i\tilde{\sfA}^{(1)}_i\right)\,,
\end{align}
where we used $\kappa\,\mathfrak{n}(\gamma)\,\kappa=\mathfrak{n}(\gamma)^{\natural}$. Since we have the relation (\ref{eq:r-l}), the factorization takes the canonical one (\ref{mm-fac}). 
In this way, we obtain the coset matrix
\begin{align}\label{coset-single}
   M(z,\rho)= -\kappa\exp\left(\sum_{i=1}^{2}\nu_i \tilde{\sfA}^{(1)}_i\right)\,.
\end{align}
Since $\nu_i=-R_i^{-1}$ is a harmonic function on the 3D base space, the coset matrix (\ref{coset-single}) gives a solution to the equations of motion for the 2D sigma model.
Furthermore, we can verify that this coset matrix is equivalent to the previous expression (\ref{ex-coset}).

\section{Multi-center extremal RN--BR black hole}

In this section, we develop a multiple-pole generalization of the monodromy matrix by exploiting the algebraic properties of the monodromy matrix constructed in the previous section. We consider this generalization so that each pole is characterized by the nilpotent matrix $\mathfrak{n}(\gamma)$, as in the previous section. Consequently, the matrices appearing in the exponent all commute with one another, and the monodromy matrix factorizes in a straightforward manner. We then show that the resulting multi-center black-hole solution is a multi-center extremal RN--BR black hole, which can be interpreted as a Majumdar–Papapetrou-type solution.

\subsection{Constructing multi-center extremal RN--BR solutions}

In the previous case, the monodromy matrix (\ref{sing-ex-RN}) is characterized by two poles located at
\begin{align}
    w_1=0\,,\qquad w_2=-\frac{1}{B}\,.
\end{align}
The first pole is associated with the extremal black-hole center, whereas the second is inherited from the Bertotti--Robinson background.  We generalize these two types of pole data to $N$ black-hole-type poles and $M$ background-type poles, respectively, with positions
\begin{align}
    w_1^{(j)}=a_j\,,\qquad  w_2^{(k)}=-\frac{1}{B_k}\,.
\end{align}
Here, $a_j$ specifies the position of the $j$-th extremal black hole on the $z$-axis.  For the second family, the positions are determined by the parameters $B_k$, which characterize the Bertotti--Robinson background. 
The corresponding modified residue matrices are chosen as
\begin{align}
   \tilde{\sfA}_{1,j}^{(1)}&=m_j\,\mathfrak{n}(\gamma)\,,\qquad
    \tilde{\sfA}_{2,k}^{(1)}=B_k^{-1}\mathfrak{n}(\gamma)
\,.
\end{align}
All residues therefore carry the same electromagnetic duality angle $\gamma$ and obey
\begin{align}[\tilde{\sfA}_{I,j}^{(1)},\tilde{\sfA}_{J,k}^{(1)}]=0\,,\qquad I,J\in\{1,2\}\,.
\end{align}
The monodromy matrix for the multi-center configuration is thus
 \begin{align}\label{multi-rn-monodromy}
 \cM(w)=-\kappa\exp\left[
 \sum_{j=1}^{N}\frac{\tilde{\sfA}_{1,j}^{(1)}}{w-w_1^{(j)}}
 +\sum_{k=1}^{M}\frac{\tilde{\sfA}_{2,k}^{(1)}}{w-w_2^{(k)}}
\right] \,.
\end{align}
Since $\mathfrak{n}(\gamma)^3=0$, the monodromy matrix can also have poles of order at most two.

\medskip

This generalized monodromy matrix is easily factorized by virtue of the mutual commutativity of the matrices appearing in the exponent.
For each pole, we introduce
\begin{align}
    \la_i^{(j)}&=\frac{1}{\rho}\left(z-w_i^{(j)}+\sqrt{(z-w_i^{(j)})^2+\rho^2}\right)\,,\\
    \nu_i^{(j)}&=-\frac{2}{\rho}\left(\la_i^{(j)}+(\la_i^{(j)})^{-1}\right)^{-1}=-\frac{1}{\sqrt{(z-w_i^{(j)})^2+\rho^2}}\,.
\end{align}
Performing the factorization as in the previous case, we obtain the coset matrix
\begin{align}\label{multi-rn-coset}
   M(z,\rho)= -\kappa\exp\left(\sum_{j=1}^{N}\nu_1^{(j)} \tilde{\sfA}^{(1)}_{1,j}+\sum_{j=1}^{M}\nu_2^{(j)} \tilde{\sfA}^{(1)}_{2,j}\right)\,.
\end{align}
Each function $\nu_i^{(j)}$ is a harmonic function on the flat three-dimensional base space satisfying 
\begin{align}
    \partial_{\rho}(\rho \partial_{\rho}\nu_i^{(j)})+\partial_{z}(\rho \partial_{z}\nu_i^{(j)})=0\,.
\end{align}
Hence, we can check that the coset matrix solves the equations of motion (\ref{sigma-eom}).

\medskip

The scalar fields obtained from the coset matrix are
\begin{align}
    e^{\phi}&=U^2\,,\qquad
    \chi_e=U^{-1}\sin\gamma\,,\qquad
    \chi_m=-U^{-1}\cos\gamma\,,\qquad
    \psi=0\,,
\end{align}
where the scalar factor $U$ is given by
\begin{align}
    U&=\sum_{j=1}^{N}m_j\nu_1^{(j)}+\sum_{j=1}^{M}\frac{1}{B_j}\nu_2^{(j)}\no\\
    &=
    -\sum_{j=1}^{N}\frac{m_j}{\sqrt{(z-a_j)^2+\rho ^2}}-\sum_{j=1}^{M}\frac{1}{B_j\sqrt{\left(z+\frac{1}{B_j}\right)^2+\rho ^2}}\,.
\end{align}
The duality relation (\ref{fomega-dual}) implies $d\omega=0$, and we then choose 
\begin{align}
    \omega=0\,.
\end{align}
The remaining Hodge-duality relation for the gauge field $\tilde{A}$ is given by
\begin{align}
   d\tilde{A}=-e^{\phi} \star_3d\chi_m\,,
\end{align}
which is equivalent to the differential equations
\begin{align}
    \partial_{\rho}\tilde{A}_{\phi}=-\rho\cos\gamma\,\partial_zU\,,\qquad  \partial_{z}\tilde{A}_{\phi}=\rho\cos\gamma\,\partial_\rho U\,.
\end{align}
The integrability condition for these equations is satisfied by the harmonicity of $U$.  Integrating these equations, we obtain
\begin{align}
    \tilde{A}_{\phi}&=\sum_{j=1}^{N}\frac{m_j (z-a_j) \cos\gamma}{\sqrt{(z-a_j)^2+\rho ^2}}+\sum_{j=1}^{M}\frac{(z+B_j^{-1}) \cos\gamma}{B_j\sqrt{\left(z+B_j^{-1}\right)^2+\rho ^2}}\,.
\end{align}
As a result, we obtained a multi-center extremal RN--BR solution
\begin{align}
ds_4^2 &= -U^{-2}dt^2 + U^2\Bigl[d\rho^2+dz^2+\rho^2d\phi^2\Bigr]\,, \\
A &= \frac{1}{U}\sin\gamma\,dt + \biggl[\sum_{j=1}^{N}\frac{m_j (z-a_j) \cos\gamma}{\sqrt{(z-a_j)^2+\rho ^2}}+\sum_{j=1}^{M}\frac{(z+B_j^{-1}) \cos\gamma}{B_j\sqrt{\left(z+B_j^{-1}\right)^2+\rho ^2}}\biggr]d\phi\,.
\end{align}
This is a Majumdar-Papapetrou-type black hole solution.

\subsection{Non-axisymmetric solution}

The monodromy construction only constructs axisymmetric gravitational solutions, but the constructed fields depend only on a harmonic function on the flat three-dimensional base.  Therefore, once the solution has been obtained, the collinear sources can be moved to arbitrary positions in $\mathbb{R}^3$.
This observation allows us to obtain the following class of non-axisymmetric solutions:
\begin{align}
U&=
-\sum_{j=1}^{N}
\frac{m_j}{\left|\mathbf{x}-\mathbf{x}_j\right|}
-\sum_{\hat{\jmath}=1}^{M}
\frac{1/B_{\hat{\jmath}}}
{\left|\mathbf{x}-\mathbf{x}^{B}_{\hat{\jmath}}\right|}\,,\\
A&=\frac{1}{U}\sin\gamma\,dt-\sum_{j=1}^{N}
\frac{m_j(z-z_j)\cos\gamma}{\left|\mathbf{x}-\mathbf{x}_j\right|}
\frac{(y-y_j)\,dx-(x-x_j)\,dy}{(x-x_j)^2+(y-y_j)^2}\no
\\
&\quad-\sum_{\hat{\jmath}=1}^{M}
\frac{(z+B_{\hat{\jmath}}^{-1})\cos\gamma}
{ B_{\hat{\jmath}} \left|\mathbf{x}-\mathbf{x}^{B}_{\hat{\jmath}}\right|}
\frac{ (y-y^{B}_{\hat{\jmath}})\,dx -(x-x^{B}_{\hat{\jmath}})\,dy
}{ (x-x^{B}_{\hat{\jmath}})^2+(y-y^{B}_{\hat{\jmath}})^2
}\,,
\end{align}
where
\begin{align}
\mathbf{x}=(x,y,z),\qquad
\mathbf{x}_j=(x_j,y_j,z_j),\qquad
\mathbf{x}^{B}_{\hat{\jmath}}
=
\bigl(
x^{B}_{\hat{\jmath}},
y^{B}_{\hat{\jmath}},
-B_{\hat{\jmath}}^{-1}
\bigr)\,.
\end{align}

\section{Israel-Wilson-Perj\'es solutions with Bertotti-Robinson asymptotics}

The nilpotent matrix $\mathfrak{n}(\gamma)$ that characterizes the monodromy matrix (\ref{multi-rn-monodromy}) associated with the extremal Reissner–Nordstr\"om black hole in the Bertotti–Robinson geometry can be further generalized.
The resulting multi-center solutions are then identified as Israel-Wilson-Perj\'es-type multi-center black hole solutions in the same geometry.

\subsection{Monodromy matrix}

To see this, we replace $\mathfrak{n}(\gamma)$ with a more general nilpotent element.
Requiring the corresponding gravitational solution to be real and $\cQ_j$ to belong to $\mathfrak{su}(2,1)$, the most general nilpotent matrix with degree 3 takes the form\footnote{From the viewpoint of solution generation, one may also consider strictly upper-triangular matrices. In this paper, however, we restrict our attention to the branch that continuously reduces to the extremal Reissner--Nordstr\"om black hole solution in the BR geometry.}
\begin{align}
 \cQ_j&=-\cN(c_j)
 =\begin{pmatrix}
 0&0&0\\
 -\sqrt{2}i\, c_j&0&0\\
 0&\sqrt{2}i\,\overline c_j&0
 \end{pmatrix}\,,
 \qquad
 \cQ_j^3=0\,.
 \label{eq:monodromy-residue}
\end{align}
where $c_i$ and $\bar{c}_i$ are arbitrary complex constant and their complex conjugates.
The generalized monodromy matrix is then given by
\begin{align}
 \cM(w)&=-\kappa\exp\left[
 \sum_{j=1}^{N}\frac{\cQ_j}{w-w_j}
 \right]\,.
 \label{eq:monodromy-exponential}
\end{align}
The matrices $\cQ_j$ do not commute in general,
\begin{align}
 [\cQ_j,\cQ_k]
 &=4i\,{\rm Im}(\overline c_jc_k)E_{31}\,,
 \label{eq:Q-commutator}
\end{align}
but the higher commutators are mutually commuting
\begin{align}
     [\cQ_i,[\cQ_j,\cQ_k]]&=0\,.
 \label{eq:Q-two-step}
\end{align}
As we will see below, the non-commutativity (\ref{eq:Q-commutator}) of the matrices $\cQ_j$ gives rise to a nontrivial source for the one-form $\omega$, and consequently the resulting multi-center black-hole solutions fall into the Israel--Wilson--Perj\'es class.

\medskip

A factorization of monodromy matrices of this type, in which the matrices $\cQ_j$ appearing in the exponent satisfy the commutation relations (\ref{eq:Q-two-step}), has already been carried out in previous work \cite{Sakamoto:2026cyo}. 
The details of the computation are given in appendix \ref{sec:fac-IWP}. Here we present only the final result.
To write down this, we introduce the complex harmonic function
\begin{align}
 \cH=\sum_{j=1}^{N}\frac{c_j}{R_j}=\sum_{j=1}^{N}\frac{{\rm Re}\,c_j}{R_j}+i\sum_{j=1}^{N}\frac{{\rm Im}\,c_j}{R_j}=H_1+i\, H_2\,,
 \label{eq:complex-harmonic}
\end{align}
where $R_j$ is the radial function around the $j$-th source at $(\rho,z)=(0,w_j)$:
\begin{align}
    R_j=\sqrt{(z-w_j)^2+\rho^2}\,.
\end{align}
The monodromy matrix is factorized into
\begin{align}
 \cM(w(\lambda,\rho,z))
 &=X_-(\lambda)M(\rho,z)X_+(\lambda)\,.
 \label{eq:canonical-factorization}
\end{align}
where the coset matrix is
\begin{align}
 M(\rho,z)=-\kappa\,e^{\cN(\cH)}\,,
 \label{eq:middle-factor}
\end{align}
and the two outer factors are given in (\ref{eq:X-def}).

\medskip

The monodromy matrix associated with the extremal Reissner–Nordström black hole in the Bertotti–Robinson geometry is recovered by the following choice of parameters. First, we assume that all complex constants $c_j$ have a common phase factor and write
\begin{align}
    c_j=q_j\,e^{i\gamma}\,,\qquad q_j\in \{-m_j,-B_j^{-1}\}\,,\qquad \gamma\in \mathbb{R}\,,
\end{align}
Furthermore, according to whether $|c_j|=m_j$ or $|c_j|=B_j^{-1}$, we choose the pole positions $\{w_j\}$ to be $a_j$ or $B_j^{-1}$, respectively. With this choice, the monodromy matrix (\ref{eq:monodromy-exponential}) reduces to the one for the extremal RN--BR black hole.

\subsection{Gravitational solutions from coset matrix}

We explicitly show that the coset matrix obtained above by factorizing the monodromy matrix solves the equations of motion of the sigma model. We then present the explicit form of the corresponding gravitational solution.

\medskip

The explicit component of the coset matrix (\ref{eq:middle-factor}) is given by
\begin{align}
 M&=-\kappa\exp\cN(\cH)=
 \begin{pmatrix}
 |\cH|^2&-\sqrt{2}i\,\overline{\cH}&1\\
 -\sqrt{2}i\,\cH&-1&0\\
 1&0&0
 \end{pmatrix}\,.
 \label{eq:coset-matrix}
\end{align}
The right-invariant current takes a strictly upper triangular matrix
\begin{align}
 d M\,M^{-1}&=
 \begin{pmatrix}
 0&\sqrt{2}i\,\,d\overline{\cH}
 &\overline{\cH}\,d\cH-\cH\,d\overline{\cH}\\
 0&0&-\sqrt{2}i\,\,d\cH\\
 0&0&0
 \end{pmatrix}\,.
 \label{eq:coset-current}
\end{align}
The coset matrix (\ref{eq:middle-factor}) can be shown to solve the sigma-model equations of motion by using the fact that $\mathcal{H}$ is a harmonic function on the 3D base space.
Moreover, the right-invariant current is strictly upper triangular, and the energy-momentum tensor vanishes,
\begin{align}
 \Tr\left[
 \partial_iMM^{-1}\partial_jMM^{-1}
 \right]&=0\,.
 \label{eq:null-current}
\end{align}
This indicates that the conformal factor $e^{2\nu}$ is trivial, and the three-dimensional base space is the flat base.

\medskip

Next, we write down the corresponding black hole solutions.
From the coset matrix (\ref{eq:coset-matrix}), the four scalar fields are obtained by
\begin{align}
 e^\phi=|\cH|^2\,,\qquad 
 \chi_e=-\frac{H_2}{|\cH|^2}\,,\qquad 
 \chi_m=-\frac{H_1}{|\cH|^2}\,,\qquad 
 \psi=0\,.
 \label{eq:iwp-scalar}
\end{align}
The Hodge dual relation becomes
\begin{align}
 \star_3d\omega&=2e^{2\phi}\left(\chi_m\,d\chi_e-\chi_e\,d\chi_m\right)\no\\
 &=2\left(H_1\,d H_2-H_2\,d H_1\right)\,.
 \label{eq:iwp-twist-complex}
\end{align}
The integrability condition follows directly from the fact that $H_{1,2}$ are harmonic functions,
\begin{align}
 d\star_3\left[2\left(H_1\,d H_2-H_2\,d H_1\right)\right]
 &=2\left(H_1\nabla^2 H_2-H_2\nabla^2 H_1\right)\text{vol}_3
 =0\,.
 \label{eq:twist-integrability-local}
\end{align}
Substituting (\ref{eq:complex-harmonic}) into (\ref{eq:iwp-twist-complex}) gives
\begin{align}
 \star_3d\omega
 =2\sum_{j<k}{\rm Im}(\overline c_jc_k) \star_3d\omega_{jk}\,,\qquad 
  \star_3d\omega_{jk}
 =R_j^{-1}d R_k^{-1}-R_{k}^{-1}d R_j^{-1}\,.
 \label{eq:twist-pair-source}
\end{align}
For the collinear configuration,
\begin{align}
 w_1<w_2<\cdots<w_N\,,
 \label{eq:ordered-axis-centres}
\end{align}
the potential $\omega_{jk}$ in (\ref{eq:twist-pair-source}) can be written as
\begin{align}
 \omega_{jk}
 =\frac{\rho^2+(z-w_j)(z-w_k)}{d_{jk}R_jR_k}d\phi\,,\qquad 
 d_{jk}=w_k-w_j>0\quad(j<k)\,.
 \label{eq:pair-potential-raw}
\end{align}
The one-form field $\omega$ is then given by
\begin{align}
 \omega&=2\sum_{j<k}{\rm Im}(\overline c_jc_k)\omega_{jk}+d\Lambda\,,
 \label{eq:omega-pair-sum}
\end{align}
where $\Lambda$ is a smooth function of $\rho$ and $z$.

\medskip

Finally, we compute the corresponding gauge potential.
To this end, we separate the spatial gauge potential $\tilde{A}$ in (\ref{ansatz}) as
\begin{align}
 \widetilde A&=\chi_e\omega+\beta\,.
 \label{eq:beta-definition}
\end{align}
Then, the associated field strength $\tilde F$ in (\ref{F-tF}) becomes
\begin{align}
 \tilde F&=d\tilde A-d\chi_e\wedge\omega
 =\chi_e\,d\omega+d\beta\,.
 \label{eq:shifted-field-beta}
\end{align}
To evaluate the Hodge duality relation (\ref{fomega-dual1}) for $\tilde{F}$, we first compute the exterior derivative of the magnetic potential $\chi_m$.
For the expression (\ref{eq:iwp-scalar}) of $\chi_m$, it gives
\begin{align}
 -|\cH|^2d\chi_m
 &=2\chi_e\left(H_1\,d H_2-H_2\,d H_1\right)-d H_1\,.
 \label{eq:maxwell-identity}
\end{align}
On the other hand, applying the Hodge dual $\star_3$ of (\ref{eq:shifted-field-beta}) and comparing it with the Hodge duality equation (\ref{fomega-dual1}), together with (\ref{eq:maxwell-identity}), we obtain
\begin{align}
 d\beta&=-\star_3d H_1\,.
 \label{eq:beta-equation}
\end{align}
For the collinear configuration, the equation (\ref{eq:beta-equation}) is solved by
\begin{align}
 \beta=\beta_\phi d\phi\,,
 \qquad
 \beta_\phi=-\sum_j{\rm Re}\,c_j\frac{z-w_j}{R_j}+\beta_0\,,
 \label{eq:beta-axisymmetric}
\end{align}
where $\beta_0$ is a real constant.

\medskip

Thus, the resulting four-dimensional solution is given by
\begin{align}
\begin{split}
 ds_4^2&=-|\cH|^{-2}(d t+\omega)^2+|\cH|^2d\bm{x}^2\,,\\
 A&=-\frac{H_2}{|\cH|^2}(d t+\omega)+\beta\,,
 \label{eq:iwp-sol}
 \end{split}
\end{align}
where $\omega$ and $\beta$ are given by (\ref{eq:omega-pair-sum}) and (\ref{eq:beta-equation}), respectively.
This multi-extremal black hole solution takes the same form as the
Israel--Wilson--Perj\'es type solutions \cite{Israel:1972vx,Perjes:1971gv}.

\subsection{Large radius behavior}

Before closing this section, we discuss the large-radius behavior of the multi-extremal black hole solution (\ref{eq:iwp-sol}). We define the total complex residue by
\begin{align}
 C&=\sum_{j=1}^{N}c_j\,.
 \label{eq:total-residue}
\end{align}
For the non-zero complex constant $C\neq0$, the large radius expansion of the harmonic function $\cH$ is
\begin{align}
 \cH=\frac{C}{r}+\frac{\bm{D}\cdot \bm{x}}{r^2}
 +\cO(r^{-3})\,,\qquad 
 \bm{D}=\sum_{j=1}^{N}c_j\bm{x}_j\,.
 \label{eq:H-asymptotic-expansion}
\end{align}
The one-form $\omega$ presented in (\ref{eq:omega-pair-sum}) does not contribute to the leading order of the metric because $\omega=\cO(r^{-2})$ in a specific gauge choice. 
Indeed, at large $r$, the potential (\ref{eq:pair-potential-raw}) satisfies
\begin{align}
 \frac{\rho^2+(z-w_j)(z-w_k)}{R_jR_k}
 &=1-\frac{d_{jk}^2\sin^2\theta}{2r^2}+\cO(r^{-3})\,.
 \label{eq:pair-asymptotic-ratio}
\end{align}
After removing the constant term by a gauge transformation, the one-form field $\omega$ falls as
\begin{align}
 \omega_\phi
 &=-\frac{\sin^2\theta}{r^2}
 \sum_{j<k}d_{jk}{\rm Im}(\overline c_jc_k)+\cO(r^{-3})\,.
 \label{eq:omega-asymptotic-falloff}
\end{align}
After performing the constant rescaling
\begin{align}
 t&=|C|^2\tau\,,
 \label{eq:br-time-rescaling}
\end{align}
the metric at the leading order becomes
\begin{align}
 ds_4^2&=|C|^2\left(
 -r^2d\tau^2+\frac{d r^2}{r^2}+d\theta^2+\sin^2\theta d\phi^2
 \right)+2\sin^2\theta\sum_{j<k} d_{jk}\,\operatorname{Im}\!\left(\bar{c}_j c_k\right)\,d\tau\,d\phi+\cO(r^{-1})\,.
 \label{eq:br-asymptotic-metric}
\end{align}
Hence, the solution (\ref{eq:iwp-sol}) with $|C|\neq 0$ describes Israel-Wilson-Perj\'es solutions with the Bertotti-Robinson asymptotics.
The field strength at large $r$ becomes
\begin{align}
    F_{\infty}={\rm Im}\,C\,d\tau \wedge d r+{\rm Re}\,C\,\sin \theta d\theta \wedge d\phi\,.
\end{align}
The asymptotic electric-magnetic charge is then expressed as
\begin{align}
 Q_\infty=-\frac{1}{4\pi}\int_{S^{2}_{\infty}}\star_4F={\rm Im}\,C\,,
 \qquad
 P_\infty=-\frac{1}{4\pi}\int_{S^{2}_{\infty}}F =-{\rm Re}\,C\,.
 \label{eq:asymptotic-charges}
\end{align}
When $C=0$, the leading harmonic term is dipolar or of higher order.  Then $r^2|\cH|^2$ has no nonzero direction-independent limit, the two-sphere does not approach a finite constant radius, and the asymptotic spacetime may not approach the Bertotti--Robinson geometry.

\section{Conclusion and discussion}\label{sec:conclusion}

In this work, we have constructed a new class of exact multi--black hole solutions immersed in the Bertotti--Robinson spacetime by employing the monodromy-matrix formalism. Starting from the extremal Reissner--Nordstr\"om black hole in the Bertotti--Robinson background, we derived the associated coset and monodromy matrices and showed that they are governed by nilpotent algebraic structures. A key observation is that the monodromy matrix for this class of solutions can be expressed in an exponential form generated by nilpotent elements of the underlying symmetry algebra. Exploiting this structure, we demonstrated that the monodromy matrix admits an explicit factorization, which allows one to reconstruct the corresponding gravitational solution in a systematic and purely algebraic manner. This construction provides a concrete realization of how integrable structures can be used to generate nontrivial gravitational solutions.
We further extended this framework to multi-center configurations by considering superpositions of poles in the monodromy matrix. The resulting solutions take a Majumdar--Papapetrou--type form, but with Bertotti--Robinson asymptotics rather than asymptotic flatness. In addition, we generalized the construction to stationary configurations and to a broader class of Israel--Wilson--Perj\'es-type solutions by allowing more general nilpotent elements in the monodromy matrix. 
In this generalized setting, the residue matrices do not necessarily commute, yet the algebraic structure remains sufficiently constrained to permit an explicit factorization via the Baker--Campbell--Hausdorff expansion. This highlights the robustness of the monodromy-matrix approach in handling more complicated configurations beyond the simplest symmetric cases.

\medskip

Finally, we examined the asymptotic structure of the solutions and showed that, for generic nonvanishing total residue, the spacetime approaches the Bertotti--Robinson geometry at large distances. This behavior contrasts with asymptotically flat multi-black hole solutions and reflects the influence of the external electromagnetic field. 
Our results therefore provide a unified framework for describing multi-black hole configurations in nontrivial backgrounds using integrable techniques. We expect that the present formalism can be further extended to more general settings, including non-extremal configurations, higher-dimensional theories, and solutions with different asymptotic structures. In particular, it would be of considerable interest to clarify the role of monodromy data in characterizing the global structure and regularity of multi-center solutions in a broader class of gravitational theories.

\medskip
\paragraph*{Note added.}
When all the point sources of the harmonic function $U$ are aligned along the $z$-axis, our solution reduces to the multi-centered solution recently derived in~\cite{DiPinto:2026rvp}, which can be interpreted as the near-horizon limit of multi--Reissner--Nordstr\"om black holes~\cite{Maldacena:1998uz}.

\section*{Acknowledgements}

J.S. was supported by the JSPS Grant-in-Aid for Transformative Research Areas (A) “Extreme Universe” No. 21H05190.

\appendix

\section*{Appendix}

\section{Matrix realization of $SU(2,1)$ and its Lie algebra $\mathfrak{su}(2,1)$}\label{sec:notation-su21}

In this appendix, we present a matrix realization of the Lie algebra $\mathfrak{su}(2,1)$ and the relevant coset space $SU(2,1)/(SL(2,R)\times U(1))$. 

\medskip

The group $SU(2,1)$ is the subgroup of $SL(3,\mathbb{C})$ that preserves a Hermitian form $\kappa$ of signature $(+,+,-)$:
\begin{align}
SU(2,1) &= \left\{g\in SL(3,\mathbb{C})\;:\; g^\dagger\kappa g=\kappa\right\}, \\
\kappa &=
\begin{pmatrix}
0&0&-1\\
0&1&0\\
-1&0&0
\end{pmatrix}\,.\label{kappa-rep}
\end{align}
The corresponding Lie algebra $\mathfrak{su}(2,1)$ consists of traceless complex matrices $x$ satisfying
\begin{align}
x^\dagger\kappa + \kappa x &= 0\,.
\end{align}
This real Lie algebra is a non-split real form of $\mathfrak{sl}(3,\mathbb{C})$.

\medskip

The Cartan--Weyl generators of $\mathfrak{sl}(3,\mathbb{C})$ are taken as
\begin{align}
\begin{split}
h_1 &= \frac{1}{\sqrt{3}}
\begin{pmatrix}
1&0&0\\
0&-2&0\\
0&0&1
\end{pmatrix}\,,
\qquad
h_2 =
\begin{pmatrix}
1&0&0\\
0&0&0\\
0&0&-1
\end{pmatrix}\,, \\
e_1 &=
\begin{pmatrix}
0&0&0\\
0&0&1\\
0&0&0
\end{pmatrix}\,,
\qquad
e_2 =
\begin{pmatrix}
0&1&0\\
0&0&0\\
0&0&0
\end{pmatrix}\,,
\qquad
e_3=[e_2,e_1]\,,
\qquad f_i=e_i^T\,.
\end{split}
\end{align}
In terms of these $\mathfrak{sl}(3,\mathbb{C})$ generators, the real span defining $\mathfrak{su}(2,1)$ is
\begin{align}
\mathfrak{su}(2,1)
&= \operatorname{span}_{\mathbb{R}}\Big\{
 i\sqrt{3}h_1,\; h_2,\; e_1+e_2,\; f_1+f_2,i(e_2-e_1),\; i(f_2-f_1),\; ie_3,\; if_3\Big\}\,.
\end{align}
The elements $i\sqrt{3}h_1$ and $h_2$ form the Cartan subalgebra in this real form. The positive generators used in the scalar parametrization are $e_1+e_2$, $i(e_2-e_1)$, and $ie_3$; their negative partners are $f_1+f_2$, $i(f_2-f_1)$, and $if_3$.

\medskip

For the timelike reduction, the relevant subalgebra is the maximally non-compact subalgebra
\begin{align}
\mathfrak{sl}(2,\mathbb{R})\oplus\mathfrak{u}(1)
&= \left\{x\in\mathfrak{su}(2,1)\;:\; x^\dagger=-\eta x\eta^{-1}\right\}\,, \\
\eta &= \operatorname{diag}(1,-1,1)\,.
\end{align}
Hence, the generalized transposition associated with this coset is
\begin{align}\label{transpose}
x^\natural &= \eta x^\dagger\eta^{-1}, \qquad x\in\mathfrak{su}(2,1)\,.
\end{align}
For the gauge generators, one has $x^\natural=-x$.

\section{Factorization of monodromy matrix}\label{sec:fac-IWP}

In this appendix, we factorize the generalized monodromy matrix (\ref{eq:monodromy-exponential}).
While the computation involved in the factorization is essentially identical, we perform it explicitly for completeness.

\medskip

For this purpose, it is useful to introduce the three matrix-valued functions
\begin{align}
 L=\sum_{j=1}^{N}\ell_j\cQ_j\,,\qquad
 C_0=\sum_{j=1}^{N}\nu_j\cQ_j\,,\qquad
 R=\sum_{j=1}^{N}r_j\cQ_j\,.
 \label{eq:LCR-definitions}
\end{align}
The matrix function $C_0$ can also be written as
\begin{align}
 C_0&=\cN(\cH)\,.
 \label{eq:C0-equals-N}
\end{align}
In this notation, the monodromy matrix (\ref{eq:monodromy-exponential}) becomes
\begin{align}\label{monodromy-IWP}
     \cM(w)&=-\kappa\exp\Bigl(L+C_0+R\Bigr)\,.
\end{align}
We factorize this monodromy matrix using the Baker--Campbell--Hausdorff formula. Following \cite{Sakamoto:2026cyo}, we verify the factorization by starting from its proposed final form.

\medskip

For this purpose, we define the functions
\begin{align}
 T_{jk}&=\ell_j\nu_k-\ell_k\nu_j
 +\ell_jr_k-\ell_kr_j
 +\nu_jr_k-\nu_kr_j\,.
 \label{eq:TAB-definition}
\end{align}
It admits a decomposition into two parts with complementary pole sets $\{\la_j\}$ and $\{-\la_j^{-1}\}$:
\begin{align}
 T_{jk}&=t^-_{jk}+t^+_{jk}\,,
 \label{eq:TAB-split}\\
 t^-_{jk}
 &=\frac{\nu_j\nu_k}{1+\lambda_j\lambda_k}
 \left(
 \frac{\lambda_j}{\lambda-\lambda_j}
 -\frac{\lambda_k}{\lambda-\lambda_k}
 \right)\,,
 \label{eq:tminus}\\
 t^+_{jk}
 &=\frac{\nu_j\nu_k}{1+\lambda_j\lambda_k}
 \left(
 \frac{1}{1+\lambda\lambda_k}
 -\frac{1}{1+\lambda\lambda_j}
 \right)\,.
 \label{eq:tplus}
\end{align}
The first term has poles only at $\lambda=\lambda_j$, whereas the second has poles only at $\lambda=-\lambda_j^{-1}$. We then define two functions valued in the commutator algebra of the nilpotent elements:
\begin{align}
 Z_-=-\frac{1}{2}\sum_{j<k}t^-_{jk}[\cQ_j,\cQ_k]\,,\qquad
 Z_+=-\frac{1}{2}\sum_{j<k}t^+_{jk}[\cQ_j,\cQ_k]\,.
 \label{eq:Zmat}
\end{align}
Because the nilpotent elements $\cQ_j$ satisfy the commutation relations (\ref{eq:Q-two-step}), the logarithm of the product of three exponentials terminates at finite order,
\begin{align}
 \log\left[
 e^{L+Z_-}e^{C_0}e^{R+Z_+}
 \right]
 &=L+C_0+R+Z_-+Z_+
 \nonumber\\
 &\quad+\frac{1}{2}\left(
 [L,C_0]+[L,R]+[C_0,R]
 \right)\,.
 \label{eq:three-factor-bch}
\end{align}
One can easily verify the identity
\begin{align}
    \frac{1}{2}\left(
 [L,C_0]+[L,R]+[C_0,R]
 \right)=\frac{1}{2}\sum_{j<k}T_{jk}[\cQ_j,\cQ_k]=-(Z_-+Z_+)\,.
\end{align}
This indicates that the monodromy matrix (\ref{monodromy-IWP}) factorizes as
\begin{align}
 \cM(w(\lambda,\rho,z))
 &=X_-(\lambda)M(\rho,z)X_+(\lambda)\,.
\end{align}
where the coset matrix is
\begin{align}
 M(\rho,z)=-\kappa\,e^{\cN(\cH)}\,,
\end{align}
and the two outer factors are
\begin{align}
 X_+(\lambda)=e^{R+Z_+}\,,\qquad X_-(\lambda)=\kappa e^{L+Z_-}\kappa\,.
 \label{eq:X-def}
\end{align}
The elementary functions satisfy
\begin{align}
 r_j(-1/\lambda)=\ell_j(\lambda)\,,\qquad 
 t^+_{jk}(-1/\lambda)=-t^-_{jk}(\lambda)\,.
 \label{eq:spectral-involution-functions}
\end{align}
Moreover, the matrix involution acts as
\begin{align}
 \kappa\cQ_j\kappa=\cQ_j^{\natural}\,,\qquad
 \kappa[\cQ_j,\cQ_k]\kappa=-[\cQ_j,\cQ_k]^{\natural}\,.
 \label{eq:residue-involution}
\end{align}
Combining \eqref{eq:spectral-involution-functions} and \eqref{eq:residue-involution} leads to the relation
\begin{align}
 X_-(\lambda)&=X_+^{\natural}(-1/\lambda)\,.
 \label{eq:BM-involution-check}
\end{align}
At $\lambda=0$, we have $r_j(0)=0$ and $t^+_{jk}(0)=0$, the function $X_+(\la)$ satisfies the boundary condition
\begin{align}
 X_+(0)=1\,.
 \label{eq:Xplus-normalization}
\end{align}
The exponent of $X_+$ has poles only in the set $\{-\lambda_j^{-1}\}$ and is analytic in a neighborhood of $\lambda=0$. The exponent of $X_-$ has the complementary pole set $\{\lambda_j\}$.  
In this way, the factorization (\ref{eq:canonical-factorization}) is precisely the desired canonical factorization.


\begin{thebibliography}{99}



\bibitem{Israel1964}
W. Israel and K. A. Khan, “Collinear particles  and Bondi dipoles in General Relativity,” Nuovo Cim. 33 (1964) 331.


\bibitem{Kramer1980}
D. Kramer, G. Neugebauer, “The superposition of two Kerr solutions,” Phys. Lett. {\bf 75}A 259 (1980).



\bibitem{Majumdar:1947eu}
S.~D.~Majumdar,
``A class of exact solutions of Einstein's field equations,''
Phys. Rev. \textbf{72}, 390-398 (1947).

\bibitem{Papapetrou}
A. Papapetrou, ``A Static solution of the equations of the gravitational field for an arbitrary charge distribution," Proc. Roy. Irish Acad. A \textbf{51}, 191 (1947).



\bibitem{Israel:1972vx}
W.~Israel and G.~A.~Wilson,
``A class of stationary electromagnetic vacuum fields,''
J. Math. Phys. \textbf{13}, 865-871 (1972).


\bibitem{Perjes:1971gv}
Z.~Perjes,
``Solutions of the coupled Einstein Maxwell equations representing the fields of spinning sources,''
Phys. Rev. Lett. \textbf{27}, 1668 (1971).

\bibitem{Teo:2023wfd}
E.~Teo and T.~Wan,
``Multicentered rotating black holes in Kaluza-Klein theory,''
Phys. Rev. D \textbf{109}, no.4, 044054 (2024)
[arXiv:2311.17730 [gr-qc]].


\bibitem{Bonnor:1954tis}
W.~B.~Bonnor,
``Static Magnetic Fields in General Relativity,''
Proc. Roy. Soc. Lond. A \textbf{67}, no.3, 225 (1954)

\bibitem{Melvin:1963qx}
M.~A.~Melvin,
``Pure magnetic and electric geons,''
Phys. Lett. \textbf{8}, 65-70 (1964)

\bibitem{Bertotti:1959pf}
B.~Bertotti,
``Uniform electromagnetic field in the theory of general relativity,''
Phys. Rev. \textbf{116}, 1331 (1959)


\bibitem{Robinson:1959ev}
I.~Robinson,
``A Solution of the Maxwell-Einstein Equations,''
Bull. Acad. Pol. Sci. Ser. Sci. Math. Astron. Phys. \textbf{7}, 351-352 (1959)

\bibitem{Ernst:1976mzr}
F.~J.~Ernst,
``Black holes in a magnetic universe,''
J. Math. Phys. \textbf{17}, no.1, 54-56 (1976)


\bibitem{Ernst:1976bsr}
F.~J.~Ernst and W.~J.~Wild,
``Kerr black holes in a magnetic universe,''
J. Math. Phys. \textbf{17}, no.2, 182 (1976)

\bibitem{Aliev:1989wz}
A.~N.~Aliev and D.~V.~Galtsov,
``Exact Solutions For Magnetized Black Holes,''
Astrophys. Space Sci. \textbf{155}, 181 (1989)


\bibitem{Alekseev:1996fq}
G.~A.~Alekseev and A.~A.~Garcia,
``Schwarzschild black hole immersed in a homogeneous electromagnetic field,''
Phys. Rev. D \textbf{53}, 1853-1867 (1996)


\bibitem{Podolsky:2025tle}
J.~Podolsky and H.~Ovcharenko,
``Kerr Black Hole in a Uniform Bertotti-Robinson Magnetic Field: An Exact Solution,''
Phys. Rev. Lett. \textbf{135}, no.18, 181401 (2025)
[arXiv:2507.05199 [gr-qc]].

\bibitem{Ovcharenko:2025cpm}
H.~Ovcharenko and J.~Podolsk{\'y},
``New class of rotating charged black holes with nonaligned electromagnetic field,''
Phys. Rev. D \textbf{112}, no.6, 064076 (2025)
[arXiv:2508.04850 [gr-qc]].

\bibitem{Astorino:2025lih}
M.~Astorino,
``Black holes in the external Bertotti-Robinson-Bonnor-Melvin electromagnetic field,''
Phys. Rev. D \textbf{112}, no.10, 104077 (2025)
[arXiv:2508.12908 [gr-qc]].


\bibitem{Alekseev:2025czq}
G.~A.~Alekseev,
``Charged black hole accelerated by spatially homogeneous electric field of Bertotti-Robinson (AdS2 x S2) space-time,''
[arXiv:2511.06082 [gr-qc]].


\bibitem{Ovcharenko:2026byw}
H.~Ovcharenko and J.~Podolsky,
``Static black holes in an external uniform electromagnetic field: Reissner-Nordstrom accelerating in Bertotti-Robinson,''
[arXiv:2602.15462 [gr-qc]].

\bibitem{Astorino:2026kuv}
M.~Astorino,
``Black holes in rotating, electromagnetic backgrounds and topological Kerr-Newman-NUT spacetimes,''
[arXiv:2604.05017 [gr-qc]].




\bibitem{Tomizawa:2025tvb}
S.~Tomizawa, J.~Sakamoto and R.~Suzuki,
``Asymmetric dyonic multicentered rotating black holes,''
Phys. Rev. D \textbf{112}, no.12, 124047 (2025)
[arXiv:2509.17583 [hep-th]].

\bibitem{Tomizawa:2026soz}
S.~Tomizawa, J.~Sakamoto and R.~Suzuki,
``Multicentered Myers-Perry black holes in five dimensions,''
Phys. Rev. D \textbf{113}, no.12, 124011 (2026)
[arXiv:2602.16243 [hep-th]].

\bibitem{Maison:1979kx}
D.~Maison,
``EHLERS-HARRISON TYPE TRANSFORMATIONS FOR JORDAN'S EXTENDED THEORY OF GRAVITATION,''
Gen. Rel. Grav. \textbf{10}, 717-723 (1979)

\bibitem{Sakamoto:2026cyo}
J.~Sakamoto and S.~Tomizawa,
``Monodromy-Matrix Description of Extremal Multi-centered Black Holes,''
[arXiv:2604.05696 [hep-th]].








\bibitem{Belinsky:1971nt}
V.~A.~Belinsky and V.~E.~Zakharov,
``Integration of the Einstein Equations by the Inverse Scattering Problem Technique and the Calculation of the Exact Soliton Solutions,''
Sov. Phys. JETP \textbf{48}, 985-994 (1978)


\bibitem{Belinsky:1979mh}
V.~A.~Belinsky and V.~E.~Zakharov,
``Stationary Gravitational Solitons with Axial Symmetry,''
Sov. Phys. JETP \textbf{50}, 1-9 (1979)




\bibitem{Hu:2026slp}
L.~Hu, R.~G.~Cai and S.~J.~Wang,
``Thermodynamics of Kerr-Bertotti-Robinson black hole,''
[arXiv:2603.18821 [gr-qc]].

\bibitem{Breitenlohner:1986um}
P.~Breitenlohner and D.~Maison,
``On the Geroch Group,''
Ann. Inst. H. Poincare Phys. Theor. \textbf{46}, 215 (1987)
MPI-PAE/PTh-70/86.

\bibitem{Chakrabarty:2014ora}
B.~Chakrabarty and A.~Virmani,
``Geroch Group Description of Black Holes,''
JHEP \textbf{11}, 068 (2014)
[arXiv:1408.0875 [hep-th]].

\bibitem{Katsimpouri:2012ky}
D.~Katsimpouri, A.~Kleinschmidt and A.~Virmani,
``Inverse Scattering and the Geroch Group,''
JHEP \textbf{02}, 011 (2013)
[arXiv:1211.3044 [hep-th]].

\bibitem{Katsimpouri:2013wka}
D.~Katsimpouri, A.~Kleinschmidt and A.~Virmani,
``An inverse scattering formalism for STU supergravity,''
JHEP \textbf{03}, 101 (2014)
[arXiv:1311.7018 [hep-th]].

\bibitem{Katsimpouri:2014ara}
D.~Katsimpouri, A.~Kleinschmidt and A.~Virmani,
``An Inverse Scattering Construction of the JMaRT Fuzzball,''
JHEP \textbf{12}, 070 (2014)
[arXiv:1409.6471 [hep-th]].


\bibitem{Sakamoto:2025jtn}
J.~Sakamoto and S.~Tomizawa,
``Building multi-BTZ black holes through Riemann-Hilbert problem,''
JHEP \textbf{12}, 049 (2025)
[arXiv:2506.19310 [hep-th]].

\bibitem{Sakamoto:2025xbq}
J.~Sakamoto and S.~Tomizawa,
``Description of non-spherical black holes in 5D Einstein gravity via the Riemann-Hilbert problem,''
JHEP \textbf{01}, 138 (2026)
[arXiv:2510.02093 [hep-th]].


\bibitem{Sakamoto:2025sjq}
J.~Sakamoto and S.~Tomizawa,
``Monodromy-matrix description of doubly rotating black rings,''
JHEP \textbf{05}, 113 (2026)
[arXiv:2511.05353 [hep-th]].



\bibitem{Harmark:2004rm}
T.~Harmark,
``Stationary and axisymmetric solutions of higher-dimensional general relativity,''
Phys. Rev. D \textbf{70}, 124002 (2004)
[arXiv:hep-th/0408141 [hep-th]].


\bibitem{Hollands:2007aj}
S.~Hollands and S.~Yazadjiev,
``Uniqueness theorem for 5-dimensional black holes with two axial Killing fields,''
Commun. Math. Phys. \textbf{283}, 749-768 (2008)
[arXiv:0707.2775 [gr-qc]].

\bibitem{Gaiotto:2007ag}
D.~Gaiotto, W.~Li and M.~Padi,
``Non-Supersymmetric Attractor Flow in Symmetric Spaces,''
JHEP \textbf{12}, 093 (2007)
[arXiv:0710.1638 [hep-th]].


\bibitem{DiPinto:2026rvp}
A.~Di Pinto and A.~Vigan{\`o},
``Supersymmetry of the static Reissner-Nordstr{\"o}m black hole in Bertotti-Robinson ($\mathrm{AdS}_2 \times \mathbb{S}^2$),''
[arXiv:2606.11101 [hep-th]].

\bibitem{Maldacena:1998uz}
J.~M.~Maldacena, J.~Michelson and A.~Strominger,
``Anti-de Sitter fragmentation,''
JHEP \textbf{02}, 011 (1999)
[arXiv:hep-th/9812073 [hep-th]].



\end{thebibliography}
\end{document}